\documentclass[
 reprint,
 amsmath,amssymb,
 aps,
]{revtex4-2}
\usepackage{graphicx}
\usepackage{dcolumn}
\usepackage{bm}
\usepackage[mathlines]{lineno}

\usepackage{amsmath}

\numberwithin{equation}{section}
\numberwithin{table}{section}
\numberwithin{figure}{section}

\preprint{AAPM/123-QED}

\usepackage{graphicx}
\usepackage{dcolumn}
\usepackage{bm}

\usepackage{siunitx}
\usepackage{comment}
\usepackage{dcolumn}
\usepackage{bm}
\usepackage{csquotes}
\MakeOuterQuote{"}
\usepackage[mathlines]{lineno}

\usepackage{amsmath}

\numberwithin{equation}{section}
\numberwithin{table}{section}
\numberwithin{figure}{section}
\begin{document}

\preprint{AAPM/123-QED}

\title{Effects of Curved Superconducting Magnets on Beam Stability in a Compact Ion Therapy Synchrotron}

\author{H.X.Q. Norman}
\email{hannah.norman@manchester.ac.uk}
 \altaffiliation{University of Melbourne}
\author{R.B. Appleby}%

\affiliation{ 
Department of Physics and Astronomy, The University of Manchester, Oxford Road, Manchester, UK\\ Cockcroft Institute, Keckwick Lane, Daresbury, Warrington, UK
}%

\author{E. Benedetto}
\affiliation{%
South East European International Institute for Sustainable Technologies, Geneva, Switzerland 
}%
\author{S. L. Sheehy}

\affiliation{%
School of Physics, The University of Melbourne, Parkville, Victoria, Australia
}%
\altaffiliation{Australian Nuclear Science and Technology Organisation, Lucas Heights, New South Wales, Australia
}
\date{\today}

\begin{abstract}
Superconducting, curved magnets can reduce accelerator footprints by producing strong fields ($>$\SI{3}{\tesla}) and reducing the total number of magnets through their capability for combined-function multipolar fields, making them an attractive choice for applications such as heavy ion therapy. There exists the problem that the effect of strongly curved harmonics and fringe fields on compact accelerator beam dynamics is not well represented: existing approaches use integrated cylindrical multipoles to describe and model the fields for beam dynamics studies, which are invalid in curved coordinate systems  
and assume individual errors cancel out over the full machine. 
In the modelling of these machines, the effect of strongly curved harmonics and fringe fields on compact accelerator beam dynamics needs to properly included. \par An alternative approach must be introduced for capturing off-axis fields in a strongly curved magnet, which may affect long-term beam stability in a compact accelerator. In this article, we investigate the impacts of deploying a curved canted-cosine-theta (CCT) superconducting magnet in a compact medical synchrotron for the first time. We develop a method to analyse and characterise the 3D curved fields of an electromagnetic model of a CCT developed for the main bending magnets of a \SI{27}{\metre} circumference carbon ion therapy synchrotron, designed within the Heavy Ion Therapy Research Integration Plus European project, \cite{HITRIplus} and the CERN Next Ion Medical Machine Study (NIMMS). The fields are modelled in the compact synchrotron in MAD-X/PTC to study their effects on beam dynamics and long-term beam stability.
The insights gained through the methods presented allow for the optimisation of both magnet and synchrotron designs, with the potential to impact the operational performance of future ion therapy facilities. 
\end{abstract}

\maketitle
\section*{Key Highlight Summary}
$\bullet$ {A tool is developed to sample field maps from strongly curved magnet models in OPERA 3D and recover multipoles and field gradients, validating the gradient calculation method.}\\
$\bullet$ {3D curved field gradients of a CCT magnet are modelled in tracking simulations of a compact medical synchrotron model for the first time.}\\
$\bullet$ {Representing field gradients as discretely sampled fields along the magnet length instead of using integrated fields drastically reduced the dynamic aperture to~25$\%$.}\\
$\bullet$ {Synchrotron can store $\epsilon_{x,y}$=\SI{5}{\text{$\mu$}\meter}, \SI{3.3}{\text{$\mu$}\meter} carbon ions for 2048 turns despite the addition the strongly curved field gradients, with the dynamic aperture recovered after tuning of higher order field gradients.}\\

\section{\label{sec:intro}Introduction}
Designs of next generation particle therapy facilities are focused on improving patient accessibility and clinical outcomes, while reducing cost and ensuring sustainability \cite{seeiist,Yap2021,norman2021reviewtechnologiesiontherapy,Farr2018NewSystems}. Much attention is given to the reduction of machine physical footprints and power consumption, while delivering high intensity beams (1$\times 10^{10}$ particles per extraction cycle) to enable efficient treatment delivery \cite{nimms,vretenar}. These critical changes will improve access to particle therapy, which could benefit around 10\% of radiotherapy patients \cite{protonstudy}, but is far less available than conventional radiotherapy \cite{PTCOG}. 
\par
One factor behind the low number of facilities is the high running and building costs associated with the accelerator and supporting beam delivery technology, 
as their physical and technical demands are too high for retrofitting into existing facilities. The accelerator alone can typically have a large footprint (e.g. \SI{75}{\meter} circumference for some medical synchrotrons \cite{Badano:385378,norman2021reviewtechnologiesiontherapy}) and high power consumption, due to dependency on normal conducting magnets. Reducing the footprint and power consumption requires magnets that exhibit both superconducting (SC) fields ($>$\SI{3}{\tesla}) and strong curvature (e.g. \SI{1.89}{\meter} bending radius for \SI{0.3}{\meter} aperture radius).\par 
Introducing curvature introduces complexity to the design and operation of the magnets and accelerator: the additional field components arising from curvature can negatively impact beam quality and stability during acceleration \cite{Goodzeit2007CombinedFM}, if they are not properly taken into account and compensated. For medical applications, acceptable limits on field quality must be determined by beam dynamics studies to comply with performance and safety measures during operation and treatment. This requires the exact field content of the magnet to be well-known for implementation in the accelerator.
\par
A caveat lies in the method of field characterisation: fields inside the magnet body are typically described in terms of 2D transverse cylindrical multipoles (Eq.~\ref{eq:2dlaplace} in Section~\ref{sec:theory}), which are integrated over the magnet length and represented as such in particle tracking codes \cite{rftrack}. A curved magnet necessitates the use of field gradients, as cylindrical multipoles 
are mathematically invalid in curved coordinate systems (see Ref.~\cite{Veres2022AMagnets}) and 
in the magnet ends, where the field is no longer translation-invariant.  
As such, integrated multipoles are insufficient to represent 3D curved magnetic fields and to gauge the full effect of curvature on the beam dynamics. 
As shown in Refs.~\cite{slac75, Veres2022AMagnets,tsoupas}, 2D and 3D field derivatives are valid for describing the fields in straight and curved coordinate systems (see Section~\ref{sec:theory}), 
 and are already specified in beam optics codes \cite{mad-x}. These provide a straightforward translation from the magnetic field content to the accelerator.
\par

The aim of this study is to understand how both the curvature and the field composition of a curved, combined-function (CF), canted-cosine-theta (CCT) coil-based magnet can impact the beam dynamics and stability in a compact medical synchrotron. While the influence of curvature has been analysed for this magnet type on the field content and quality \cite{qinbin} and on the beam dynamics in a compact ring \cite{cctisoldering}, the two aspects have not been studied together in the context of a medical synchrotron. 
To explore these effects, we develop general tools (further detailed in Sections~{\ref{sec:theory}, \ref{sec:tool}) that perform field sampling and gradient calculations for implementation in the accelerator.
\par
We investigate the suitability of a strongly curved (\SI{90}{\degree}), CCT coil-dominated magnet with CF quadrupole layers for the bending sections in a compact medical synchrotron: a
project within the HITRI-plus initiative \cite{HITRIplus},
which has produced novel accelerator designs for a planned ion therapy facility in South Eastern Europe \cite{BenedettoCompletedMagnets}. 
The strong fields (\SI{3.5}{\tesla}) and CF layers attributed to the unique geometry of the CCT (detailed in Section~\ref{sec:magnets}) enable reduction of the accelerator circumference to \SI{27}{\metre}. \par Our method
employs finite element software \cite{OperaSystemes} to simulate a model of the magnet for field analysis. 
We develop tools (\cite{Norman:2845833, hannahipac23}) to analyse and characterise the fields using gradients for direct implement by beam optics codes \cite{mad-x,Schmidt:573082}, using which we simulate particle tracking in the synchrotron lattice. From the results, we set limits on the higher order gradients 
that would otherwise drive strong resonances, leading to instabilities in the ring over more turns. Subsequently the suitability of the accelerator design for future facilities is determined.

\par{}

\section{Magnet Models}\label{sec:magnets}
A CCT configuration has been established as an attractive solution for a compact medical accelerator \cite{BenedettoCompletedMagnets,qinbin,rossisig}, but is yet to be implemented in operation. As such, two SC coil-dominated models are presented for the main bending magnets in the NIMMS SC compact synchrotron: a combined-function (dipole and quadrupole) \SI{30}{\degree} cosine-theta (CT) magnet and a \SI{90}{\degree} CCT dipole magnet with alternating-gradient focusing-defocusing (FD) quadrupole layers (Fig~\ref{fig:cctctmodel}). The AG-CCT magnet has been customised for NIMMS by LBNL, based on a similar design for a compact SC proton therapy gantry \cite{Wan2015AlternatinggradientCC}; the CT is an alternative option in case of engineering complications \cite{elena}. The specifications for both magnets are in Table~\ref{tab:cctparams}. 
\begin{figure}[!b]
\begin{center}
\begin{minipage}[b]{0.50\linewidth}
\includegraphics[width=\linewidth]{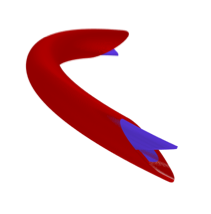}

\end{minipage}
\begin{minipage}[b]{0.48\linewidth}
\includegraphics[width=\linewidth]{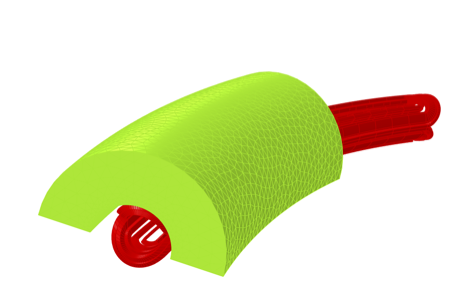}

\end{minipage}
\caption{Left: Alternating gradient canted-cosine-theta magnet with \SI{90}{\degree} bend. Right: Combined-function Cosine-theta magnet with\SI{30}{
\degree} bend. Models courtesy of L. Brouwer (LBNL) \& M. Karpinnen (CERN). 
} 
\label{fig:cctctmodel}
\end{center}
\end{figure}
\begin{table}[!hbt]
\caption{Parameters for the \SI{90}{\degree} AG-CCT and \SI{30}{\degree} Combined-Function CT}
\begin{ruledtabular}
\begin{tabular}{lll}

\centering
         
       Parameter & AG-CCT & CF-CT\\
           Main Bending Field B$_\mathrm{dip}$ [T]        & 3.5 & 3.0 \\
           Max Quadrupole Gradient [T/m] & $\pm$10 & $\pm$3.5 \\
        Magnet Physical Bend [\SI{}{\degree} & 90 & 30       \\ 
          Bending Radius [m]   &  1.89 & 2.21     \\ 
          Half Aperture Diameter ($A_{x,y}$) [mm] & 30 & 11.1 
          \\

    \end{tabular}
   
   \label{tab:cctparams}
\end{ruledtabular}
\end{table}

\par

The magnet geometries are chosen due to the simplicity of design, engineering and `tuning' (adjustment of higher order field gradients) \cite{Caspi2007DesignSolenoids}. Additional advantages include good field quality($\frac{\Delta B}{B} = 10E^{-4}$) 
and lower power consumption compared with normal conducting magnets. Both geometries produce pure cosine-theta dipole fields. CT magnets are formed from several turns of one coil; CCT geometries are formed of two (or more) coil layers, superposed with opposite directions of tilt, as shown in Fig.~\ref{fig:cct}. 
\begin{figure}[!h] 
\begin{center}
\includegraphics[width=.8\linewidth]{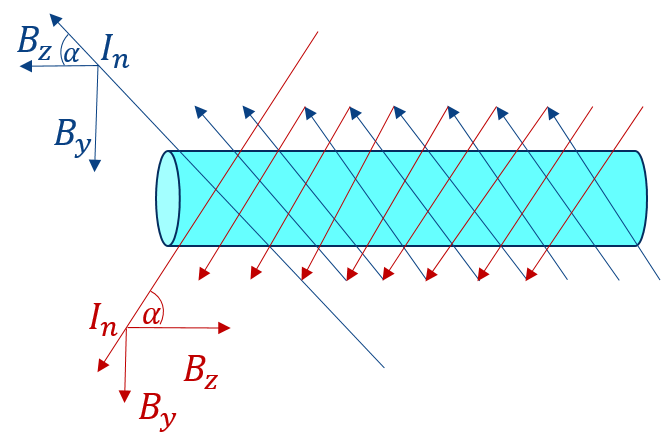} 
\caption{The current $I_n$ flows on two coil layers tilted at angle $\alpha$ from the horizontal plane $z$ and wound around a mandrel (a central body with grooves). 
$I_n$ flows in opposite directions, producing an overall main field stored in $B_y$ in the vertical plane and cancelling $B_z$ in the $z$-plane.}
\label{fig:cct}
\end{center}
\end{figure}
\par

The reasons for choosing a CCT geometry over the more common CT geometry are due to ease of tunability and stress-management. A CCT has an internal structure consisting of a mandrel (the magnet body in Fig.~\ref{fig:cct}), where the Lorentz forces are intercepted by its ribs and transferred to the spar (the centre of the magnet- see Fig.~\ref{fig:crosscct}).
\begin{figure}
\begin{center}
\includegraphics[width=0.6\linewidth]{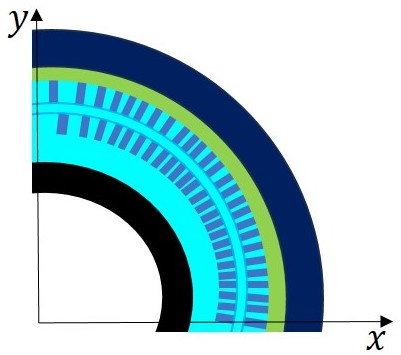} 
\caption{A quarter cross-section of a CCT coil. The conductor is shown in light blue (the widest layer in the diagram), with intercepting ribs shown (darker blue). The ribs intercept Lorentz forces which get transferred to the spar (the black innermost layer). The outermost layer is the iron yoke, in navy.} 
\label{fig:crosscct}
\end{center}
\end{figure}
As such, no pre-stress is required for a CCT magnet, unlike the CT \cite{Caspi2007DesignSolenoids}. For tunability, the winding pattern of the coil layers in a CCT can be modified to produce different orders of fields (e.g. dipole, quadrupole, including combinations, as in the AG-CCT), and to reduce unwanted higher order field gradients. Generally the field gradients of any magnet must be well-known to determine that the field quality is within design tolerances. This is important for beam dynamics studies: unwanted higher orders can give rise to instabilities in the accelerator over time, leading to beam loss. In our case, due to the compactness of the magnets and synchrotron, the field gradients must be known to high accuracy: especially in the AG-CCT 
where the FD layers overlap, and where end fringe fields will have significant effect \cite{Russenschuck:1221810}. In Section~\ref{sec:theory}, we detail the method for calculating higher order field gradients on and off the magnet midplane. 
\section{Magnetic Field Theory of Straight and Curved Magnets}\label{sec:theory}
In practice, the field vector is measured within the aperture of a simulated or physical magnet, where there is no current \cite{Jain:1997}. A 2D representation of the field can accurately describe the field over the magnet body away from the ends. To know the field distributed over an entire (coil-based) magnet, we examine the magnetic scalar potential in cylindrical polar coordinates, $V(r,\theta, z)$\footnote{It is worth pointing out at this stage that for the purpose of illustrating equations of interest, we use the general form of the 2D cylindrical Laplace equation in terms of a scalar potential, V($r,\theta$). Later we will show solutions to the Laplace equation based on a vector potential, \textbf{A}($r,\theta$) as we will need to consider additional transverse components that arise from the magnet curvature.}. 
\subsubsection*{$\bullet$ Case 1: 2D Straight System}
To find the potential everywhere within the magnet aperture, we need to solve the 2D cylindrical Laplace equation subject to boundary conditions defined for our particular system (see Appendix \ref{app:boundary}). The general form of the Laplace equation in cylindrical coordinates, in the absence of current and longitudinal variation, is 
\begin{equation}\label{eq:2dlaplace}
\nabla^{2}V(r,\theta)=\frac{\partial ^2V}{\partial r^2}+\frac{1}{r}\frac{\partial V}{\partial r}+\frac{1}{r^2}\frac{\partial ^2V}{\partial \theta^2} = 0
\end{equation} 
$r$, $\theta$ are the radial distance and subtended angle respectively, measured from the nominal beam trajectory where we evaluate the field, depicted in Fig.~\ref{fig:coord}. From Eq.\ref{eq:boundary}, we require a finite potential as $r\rightarrow$ 0. Using separation of variables, the solution to Eq.~\ref{eq:2dlaplace} 
is given by a linear combination of the basis functions, known as cylindrical multipoles, where the radial and azimuthal components of the magnetic field \textbf{B}($r,\theta$) follow as 
\begin{equation}\label{eq:multipole_r}
{B_{r,\theta}(r,\theta}) = \sum_{n=1}^{\infty} \left(\frac{r}{r_{0}}\right)^{(n-1)} \left[B_{n}\sin{n\theta} \pm A_n\cos{n\theta}\right] 
\end{equation}
where the integer $n$ = 1,2,... is the multipole order (e.g. $n$ = 1 for a dipole in European convention) \footnote{NB: this paper uses the European convention for calculating the corresponding equations for the radial (B$_r$) and the azimuthal (B$_\theta$) components of the field (the summation starts from $n$ = 1 instead of $n$ = 0).}. The arbitrary reference radius $r_{0}$ is typically chosen as $\approx\frac{2}{3}$ the coil radius: the 
upper limit within the good field region of the magnet \cite{Jain:1997}. $B_n$ and $A_n$ are the \textit{normal} and \textit{skew} multipole coefficients respectively and can be recovered by a Discrete Fourier Transform (DFT) performed on Eq.~\ref{eq:multipole_r} \cite{Russenschuck:1221810}. Field harmonics are given in terms of the main field expected in the magnet and in `units' scaled by $10^{4}$. Typically magnet designers aim for $\approx 0.1$ to $2$ units to achieve good field quality \cite{9723617}. \par 
In a 2D straight system (Cartesian or Polar), multipole expansion is equivalent to a 2D Taylor expansion on the main field along the midplane. It is sufficient to take the derivatives of the main field along the midplane (see Ref.~\cite{slac75}),
\begin{equation}\label{eq:deriv}
\frac{\partial^{n} B_y(x,y)}{\partial x^{n}}\mid_{y=0}
\end{equation}
The equivalence between the two field expansions in different systems is discussed further in Appendix \ref{app:altsoln}; it does not hold in a curved system, as we will demonstrate in Section~\ref{sec:tool}. 
\subsubsection*{$\bullet$ Case 2 : 2D Curved System}
\noindent
When considering the beam dynamics in an accelerator, we should be in the local (beam) coordinate system, known as curvilinear coordinates, shown in Fig.~\ref{fig:coord} alongside previously referred to coordinate systems.
\begin{figure}[!h]
\begin{center}
\includegraphics[width=.7\linewidth]{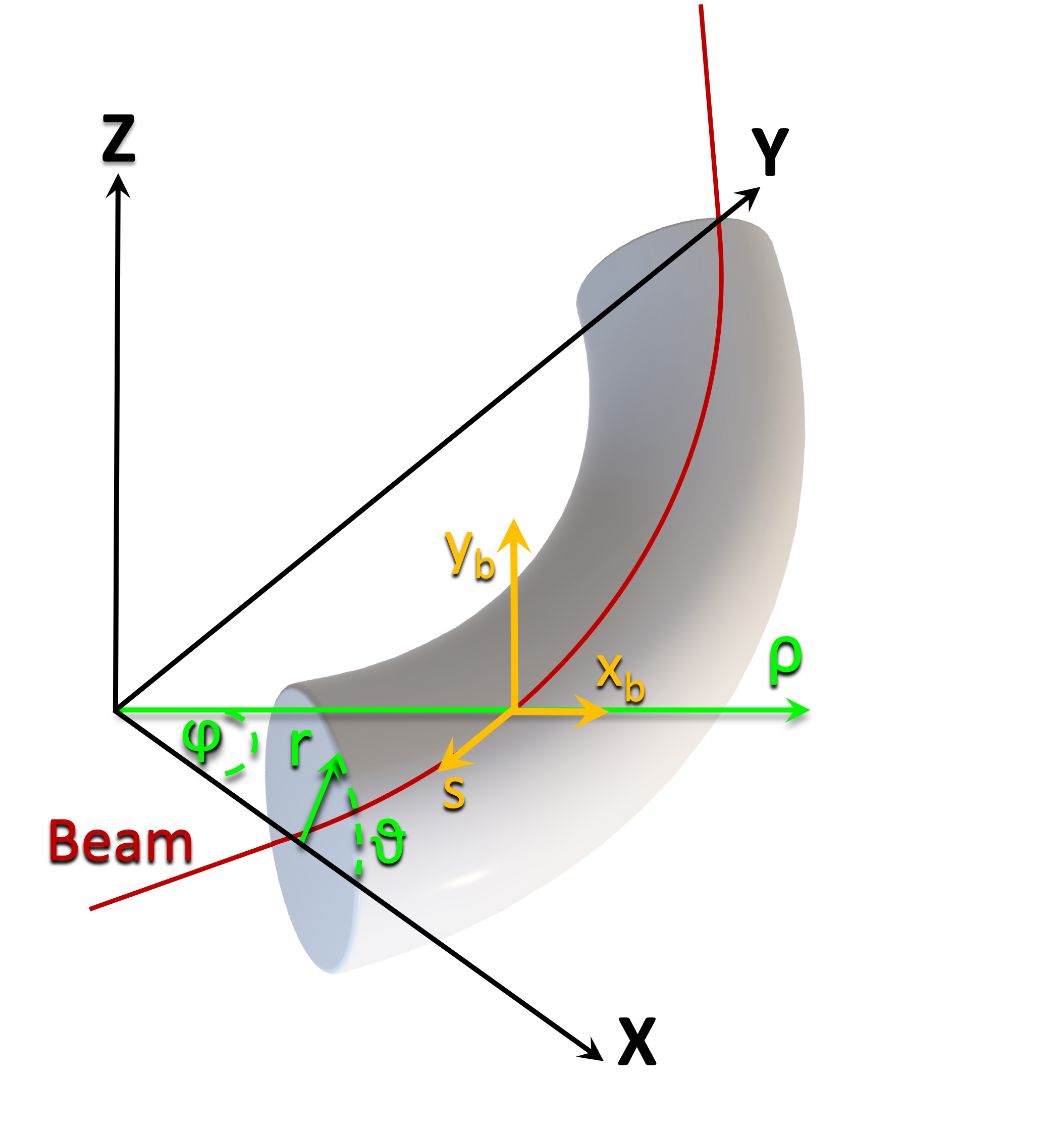} 
\caption{A curved magnet shown with straight Cartesian coordinates ($x,y,z$) and curvilinear Cartesian coordinates ($x_{b},y_{b},s$) (the local coordinate system experienced by a beam travelling through the magnetic field). The magnet rotates around the beam by angle $\theta$. $\rho$ is the bending radius of the magnet while $r$ is the radial distance measured from the beam. $\phi$ is the angle between the global Cartesian coordinate $x$ and the local (beam) coordinate $x_b$.}
\label{fig:coord}
\end{center}
\vspace*{-1.0\baselineskip} 
\end{figure}
Now we present the solutions to the Laplace equation based on a vector potential, \textbf{A}($r,\theta$) as we consider the additional transverse components that arise from magnet curvature. Switching to 2D curvilinear coordinates and satisfying the Coulomb gauge prescription  ($\nabla \cdot \textbf{A} = 0$) results in the general form for the vector Laplacian \cite{Lee}
\begin{equation}\label{eq:curvilaplace}
\begin{split}
\nabla^{2}\textbf{A}(x,y) & = \frac{\partial ^{2}A_s}{\partial x^{2}} + \frac{\partial A_s}{\partial y^{2}} + \frac{1}{(\rho + x)}\frac{\partial A_s}{\partial x} - \frac{A_s}{(\rho + x)^{2}} 
\\ & = 0
\end{split}
\end{equation}
We can assume the vector potential to act only along the beam coordinate axis $s$; the bending radius $\rho$ is along the length of the magnet and transverse coordinates ($x,y$) are centered on the nominal trajectory. Eq.~\ref{eq:curvilaplace} is not solvable in closed form, due to added constants that depend on $x$: 
there is no linear combination of cylindrical multipoles that can form solutions for a curved magnetic field.
\par We can however find a solution \cite{ipac03, tsoupas} in the form of a power series and obtain recursion relations between the coefficients for a more generalised case (i.e. off the magnet midplane). The solution for the magnetic field, given by the solution to the vector potential that satisfies \ref{eq:curvilaplace} is 
\begin{equation}\label{eq:2dpoly}
\begin{split}
B_i(j,k,x_{i},y_{l}) &= B_{i}(j,k,0,0) + 
\sum_{j=1}\sum_{k=1}c_{i,j,k,n,m}(x_{l})^{n}(y_{l})^{m} \\
&n,m = 0,1,2,\ldots, \quad i=(1,2,3)=(x,y,z)
\end{split}
\end{equation}
Where $c_{i,n,m,j,k}$ represent the (unnormalised) field derivatives, and are given by recursion relations in \cite{slac75}. A smaller 2D `slice' of the full (global) fieldmap is characterised by the $(j,k)$ indices and the individual slice's local coordinates ($x_l,y_l$). The components of \textbf{B} are expanded along the magnet midplane in Appendix~\ref{app:altsoln}.
Conveniently, on each individual slice ($j,k$), we have a 1:1 mapping between the coefficients of \ref{eq:2dpoly} $c_{i,n,m}$ of the expansion and the field derivatives, via: \begin{equation}\label{eq:relationship1}
        c_{i,n,m} = \frac{1}{(n,m)!}\frac{d^{n,m}B_{y}}{d{x^{n,m}}}
    \end{equation}
Even in the ideal case of being on the midplane, the coefficients $c_{i,j,k,n,m}$ now have dependency on the curvature, $\rho$. Away from the midplane, we cannot assume $A_{x} = A_{y}$ = 0, therefore our expansion must be performed on \textbf{A} and not just a single scalar component of \textbf{A}. The advantage of series expansion is that the solution stays the same; whether there is variation in \textbf{A} along $s$, if we are on or off the magnet midplane. We need only to take care when including the dependence on magnet curvature when solving for the coefficients $c_{i,j,k,n,m}$ \footnote{In a 2D curved system, the terms of the field expansion along the mid-plane is now attached to both $x$ and $y$: one can relate the terms of the 2D polynomial expansion to both $B_y$ and $B_x$, therefore one should be careful to define the skew components as well as the normal components for use in tracking codes}. We see shortly how the generalised 2D polynomial expansion is applicable in 3D systems.

\subsubsection*{$\bullet$ Case 3 : 3D 
Curved System}

In close proximity to the magnet's physical ends, there are fringe fields which continue to deflect the beam beyond the ends of the coil \cite{Marks2013}. The field here is 3D, meaning the usual 2D multipole expansion is again no longer valid. There exist a few different approaches to treat the dependence of the field components on the magnet length (see \cite{Wolski_2018}). The most straightforward approach we have taken is to assume that at any one sampled measurement of a magnetic field, the beam will be subject to a 2D `slice' of the magnetic field and will have one $z$-coordinate; the problem is reduced to 2D spatially. 
\par
By sampling all three components of the field along ($x,y$) (one 2D slice) for one particular $z$ and fitting \ref{eq:2dpoly} to the main field component over the slice, we can step along in $z$ and repeat the fit over each 2D slice defined along the magnet length or region of interest, integrating the gradients at the end of the region.\par We have developed a Python tool based on this idea that samples and analyses 3D fields from \textbf{any} magnetic field map, such that the normal and skew field derivatives may be recovered at any point within the magnet. Further detail and application follows in Section~\ref{sec:tool}. 

\section{Field Sampling and Analysis (Methods)}\label{sec:tool}
Reliable methods must be available for calculating both multipole coefficients and field gradients for a given magnetic field, to compare and therefore justify using gradients to describe curved fields. In this section, we construct a Python tool that makes use of these methods. Outside of this study, we first tested the tool on a 2D, arbitrary (or `ideal') straight Cartesian magnet, which is given initial inputs to generate dipole, quadrupole, and higher order fields. Ideal fields are used such that the multipoles and gradients after calculation can be compared against the input fields. The tool was then applied to a more realistic magnet, the straight CT OPERA 3D model, to confirm the matching of field gradients and multipoles after sampling the main field in the magnet body. Using the curved CT, we sample the field and calculate both the gradients and multipoles to demonstrate the mismatch between the two quantities and to reiterate the importance of using field gradients to describe curved fields. Finally, we extend the tool to 3D for off-axis field gradient calculation.
\subsection{Field Component Calculation in 2D Curved Systems}\label{section:curved} 
We apply the tool to the magnetic fieldmap of a curved magnet, to show that multipoles are not equivalent to gradients in a curved coordinate system, and for later performing beam dynamics simulations. We define an ideal path taken by a particle through the centre of the magnet, and construct vectors perpendicular to the magnet curvature (Fig.~\ref{fig:curvedsampleline}).
\begin{figure}[!h]
\begin{center}
\begin{minipage}[!h]{0.7\linewidth}
\includegraphics[width=\linewidth]{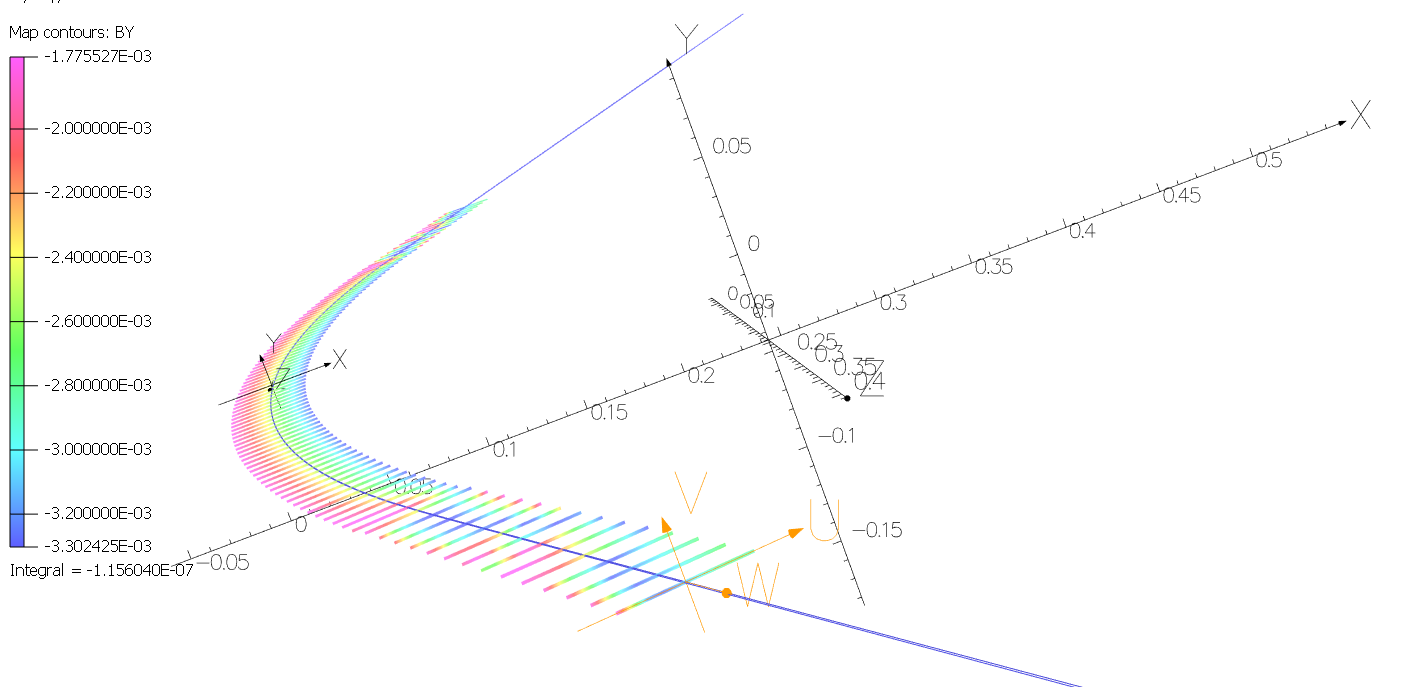}
\label{fig:curvedsampleline}
\end{minipage}
\begin{minipage}[!h]{0.7\linewidth}
\includegraphics[width=\linewidth]{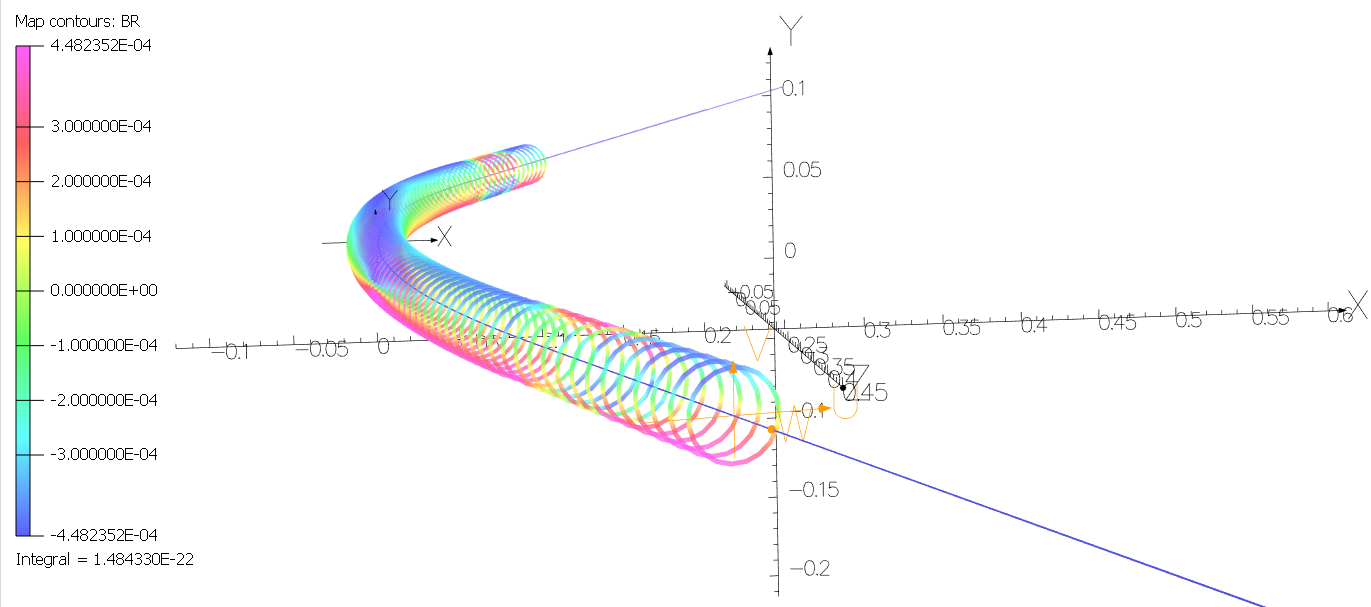}
\label{fig:curvedsamplecircle}
\end{minipage}
\caption{Main field (represented by colour scale) sampled on a nominal trajectory through a curved CT magnetic field. Top: $B_y$ sampled along $y$=0 perpendicular to each point on the path.  Bottom: $B_r $ sampled on circles defined by $r_0$ = \SI{0.02}{\meter} (2/3 the aperture radius), perpendicular to each point on the path. Local coordinate systems are defined for both sampling depictions.}
\label{fig:sampling}
\end{center}
\end{figure}
\begin{figure}[!h]
\begin{center}
\includegraphics[width=\linewidth]{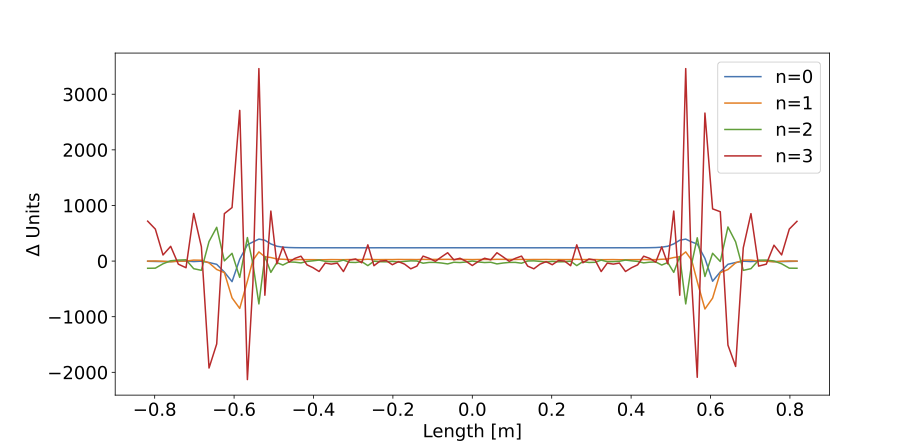}
\caption{Magnetic field sampled over the entire length of a \SI{30}{\degree} CT magnet with CF dipole/quadrupole fields ($z$=-0.8 to 0.8). Difference between 2D curved integrated gradient harmonics and integrated multipole harmonics (calculated up to $n=3$) in units $\times 10^{-4}$.}
\label{fig:2dcurvedcomp}
\end{center}
\end{figure}From Fig.~\ref{fig:2dcurvedcomp}, it is seen that the gradients and multipoles mostly agree within the body region. However, there is a stark difference between the integrated gradients and multipoles at the ends of the curved magnet. While the multipoles are seen to fall to zero at the ends, the polynomial fitting shows that this is not the case. Table \ref{tab:curvedcomp} shows the divergence between the two quantities integrated over the whole length of the magnet, emphasising the point that these two representations are not equivalent in curved coordinate systems: we must use Taylor expansion instead of Fourier analysis to calculate the magnetic field gradients in a curved system, and where there is longitudinal dependence.

\subsection{3D Field Component Calculation}\label{section:3d} 
In practice, higher order field gradients contribute to non-linear behaviour in the beam dynamics of the ring. Especially in the case of a strongly curved magnet, due to the geometry we must take transverse field components ($B_{x}$, $B_{y}$) into account, as well as the longitudinal path ($z$) and field ($B_z$). \par To calculate the strongly curved field gradients from the AG-CCT, we sample the magnet's fieldmap. To account for the curvature and for the off-axis field contributions, we fit a 2D version of the polynomial (Eq.~\ref{eq:2dpoly}) to the main field over several `slices' at one point in $z$, as shown in Fig.~\ref{fig:2Dfit}. Using the 2D polynomial we can calculate the 2D field gradients that include transverse field dependence. By using the gradients we can reproduce the original field and assess the accuracy of the fit, both graphically (Fig.~\ref{fig:2Dfit}) and analytically, as presented in Table ~\ref{tab:3dcurvedgrads}. 
\begin{figure}[!h] 
\begin{center}
\includegraphics[width=.9\linewidth]{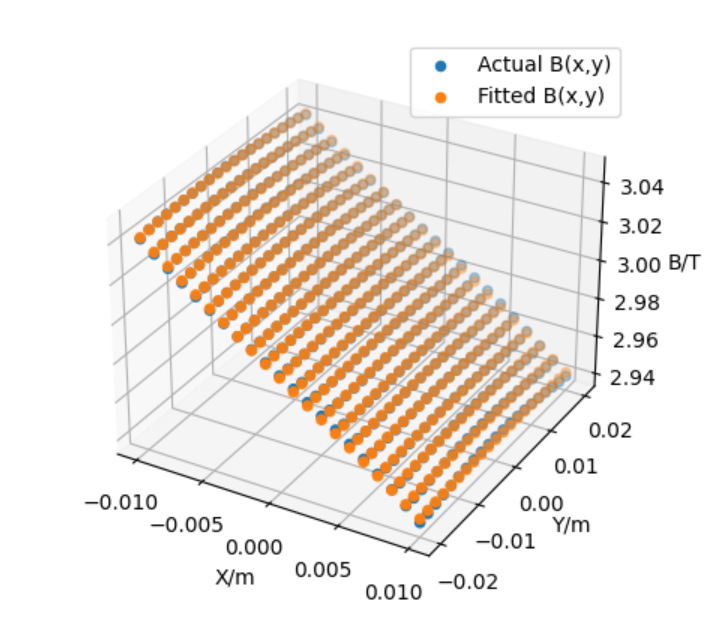} 
\caption{A comparison between the main magnetic field measured over a 2D slice of the AG-CCT fieldmap and the field reproduced by a 2D polynomial fitted to the data points.}
\label{fig:2Dfit}
\end{center}
\end{figure}

\begin{table}[!htb]
\centering
\caption{3D Curved System: Integrated gradients converted from the coefficients of a 2D polynomial fitted to main field $B_{z}$ for orders of $n =$(0,...,3), between $z$ = -1.5 and \SI{1.5}{\meter}. The integrated gradients have been scaled by $r_0$ = $2/3 * r = $\SI{50}{\milli\meter}. Each integrated gradient harmonic is calculated via $b_n$ = $B_{n}$/$B_\text{main} \times10^{4}$.}
\begin{ruledtabular}
\begin{tabular}{llll}

Order(n) & Normal $b_{n}(x,y)$ & Skew $a_{n}(x,y)$                  & Unit        \\ \hline
0        & ${3.5}$            & ${-0.002} $                         & T               \\ 
1        & ${2.79}$          & ${-0.15}$                 & $10^{-4}$                \\ 
2        & ${14.38}$               & ${-0.10}$             & $10^{-4}$    \\ 
3        & ${28.29}$              & ${-7.58}$ & $10^{-4}$   \\ 
\end{tabular}
\end{ruledtabular}
\label{tab:3dcurvedgrads}
\end{table} \par
The dipole and quadrupole gradients ($b_{0},b_{1}$) approximately match their expected design values of \SI{3.5}{\tesla}, \SI{10}{\tesla\meter^{-1}} ($b_{1}$ translated to \SI{2.79} units after integrating throughout the magnet), however the sextupole and octupole harmonics ($b_{2},b_{3}$) are outside the usual design limits of 0.1 - 2 units. As a means of verification of the 3D gradient calculation, we also applied the calculation to the curved and straight CT models and recovered the same results as calculated in Tables.~\ref{tab:curvedcomp},~\ref{tab:integratedbody}. The large higher order gradients are likely due to the high curvature and additional combined-function layers, and will need tuning in the next iteration of the design. It is evident from the results that there are also non-zero skew multipoles ($a_n$) present due to the magnet curvature, as earlier established in Section~\ref{sec:theory}). \par
In summary, we have developed a Python tool that is able to calculate field gradients and multipole coefficients in straight and curved systems, and have shown that the two representations are not equivalent in curved systems. Having established that a more accurate representation of a curved magnetic field is to use field gradients using polynomial fitting, we have defined both the normal and skew 3D gradients of the AG-CCT magnet in Table \ref{tab:3dcurvedgrads}. These results are taken forward to the next phase of the study for particle tracking simulations in MAD-X/PTC (Section~\ref{sec:beamdynamics}).

\section{Beam Dynamics Studies (Methods)}\label{sec:beamdynamics}
In this section we implement the curved magnetic field gradients calculated from the AG-CCT to study the effects in a model of a superconducting, compact carbon ion therapy synchrotron \cite{BenedettoCompletedMagnets}, depicted in Fig.~\ref{fig:nimms}, with details given earlier in Sections~\ref{sec:intro}, \ref{sec:magnets}. 
The original working point ($q_x$ = 1.68, $q_y$ = 1.13) is near a third order resonance for extraction \cite{BenedettoCompletedMagnets}. For our study, the working point is moved to ($q_x$ = 1.71, $q_y$=1.085), away from strong coupling resonances in order to examine tune shift due to field gradients. Other parameters of the synchrotron operation and tracking are given in Table~\ref{tab:synch}.  
\begin{figure}[!h] 
\begin{center}
\includegraphics[width=.75\linewidth]{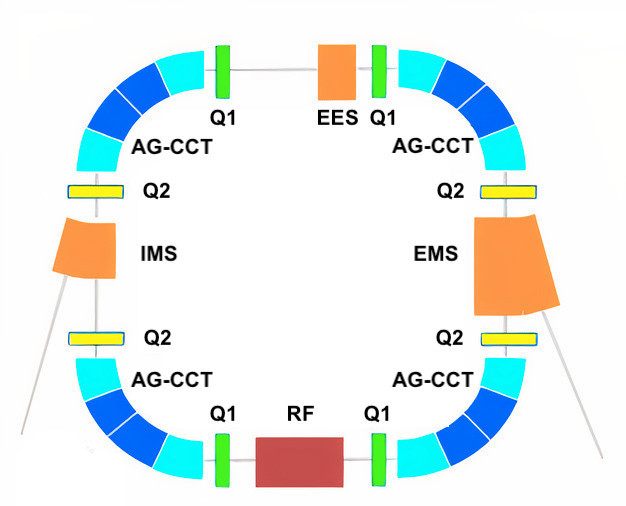} 
\caption{The NIMMS \SI{27}{\metre} superconducting synchrotron, based on four alternating-gradient CCT combined-function dipole/quadrupoles with \SI{90}{\degree} bends.}
\label{fig:nimms}
\end{center}
\end{figure}

\subsection{Optical Thin Lens Element Approximation}
The synchrotron is represented in MAD-X \cite{mad-x} as a lattice composed of four periodic cells, each containing an AG-CCT magnet (defined using the optical `sector bend' element). Particle tracking is performed using the PTC \cite{Schmidt:573082} module within MAD-X.
\begin{figure}[!h] 
\begin{center}
\includegraphics[width=0.8\linewidth]{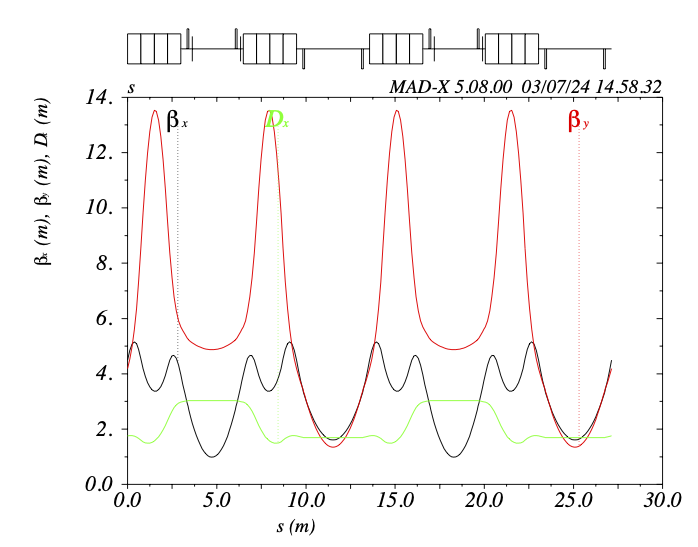} 
\caption{Optics functions $\beta_{x,y}$ and $D_x$, using magnetic elements within the NIMMS synchrotron lattice}
\label{fig:survey}
\vspace*{-1.0\baselineskip} 

\end{center}
\end{figure}
\begin{table}[!htb]
\caption{Operational and particle tracking parameters for the \SI{27}{\meter} NIMMS Synchrotron ($\beta_\text{rel} = 0.73,\gamma_\text{rel} = 1.46$).}
\begin{tabular}{lll}
 Parameter & Unit & Value \\
\hline
\\
$E_{k_\text{max}}$  & MeV/u & 430 
\\
N$_\text{particles}$ & - & $2\times{10^{10}}$
\\
 $\varepsilon_{x_N}$, $\varepsilon_{y_N}$ & $\mu$m & (0.7, 1.0)
\\
$\beta_{x,\text{max}}$, $\beta_{y,\text{max}}$ & m & (5.16, 13.53)
\\
Dispersion $D_x,\text{max}$ & m & 3.03\\
$\sigma_{x}$, $\sigma_{y}$ & mm & (0.85, 1.78)
\\
$Q_{x}$, $Q_{y}$(extraction) & - & (1.68, 1.13)
\\
$Q_{x}$, $Q_{y}$(working conditions) & - & (1.71, 1.085)\\
Natural Chromaticity $\xi_{x}$, $\xi_{y}$ & - & (-0.263, -0.207) \\ \hline
\end{tabular}
\vspace*{-1.0\baselineskip} 

\label{tab:synch}
\end{table}
Approximating the behaviour of the beam under influence of `kicks' (magnets) is dependent on how the field gradients are defined and positioned in the lattice. As earlier discussed, a common technique in accelerator physics is to integrate the gradients over an entire magnet and divide by an effective length \cite{latina}. As the AG-CCT is short ($\approx$ \SI{3}{\metre} in length) compared to the fringe field extent \cite{Hull2015} and the field gradients near the magnet ends fluctuate and do not fall to zero (as seen in Fig.~\ref{fig:2dcurvedcomp}), a more detailed representation is required. \par The integrated 3D normal and skew gradients up to $3^\text{rd}$ order (octupole) of the AG-CCT magnet (in Table~\ref{tab:3dcurvedgrads}) are represented in MAD-X using discretely sampled fields from each point in magnet length for performing beam dynamics simulations. The field gradients are represented as many thin lenses ($>60$), interleaved with sub-divided sector-bends within the magnet. The tracking results comparing both representations (see Section~\ref{sec:results}) differ significantly, justifying our choice.
\subsection{Methods: Beam Dynamics Simulations (Stability Studies)} 
\subsubsection*{Linear Dynamics Tools: Phase Space and Tune Shift with Action} 
The $\beta_{x,y}$ and $D_x$ optical functions with respect to the synchrotron lattice elements are shown in Fig.~\ref{fig:survey}. Initial ($x,y$) coordinates are scanned up to the magnet half-aperture ($A_{x,y}$=\SI{30}{\milli\metre}) \cite{BenedettoCompletedMagnets}, which accommodates an initial amplitude offset of up to $10\sigma_{x,y}$ (see Table~\ref{tab:synch}). On-momentum particles are launched in $x$ from 0 -- $10\sigma_x$ and scanned separately in $y$ from 0 -- $10\sigma_y$. The normalised horizontal ($x,x'$) and vertical ($y,y'$) phase space is then analysed as a first approach to identify any beam distortion arising from the presence of the gradients. 
\par

To gain insight into the effects of the gradients on long term beam stability, the amplitude-dependent tune shift is examined. Using the same tracking data, fractional tunes $q_{x,y}$ are calculated for each initial $x,y$ amplitudes, using the NAFF algorithm \cite{GitHubPyCOMPLETE/NAFFlib} on the positions of each particle over many turns. We use invariant action-angle variables ($J_{x,y}, \phi_{x,y}$), to examine the change in tune of the particles from the initial starting amplitudes \cite{Lee}. 
\subsubsection*{Non-Linear Dynamics Tools: Frequency Map Analysis and Dynamic Aperture}
We examine the non-linear effects of the curved CCT field gradients on the beam, such that the overall performance of the synchrotron can be determined subject to the presence of curved gradients. Frequency Map Analysis (FMA) is used to study the relationship between initial particle amplitudes and their resulting tunes after 2048 turns to estimate the tune spread and stability of the beam over time \cite{laskar}. This is viewed on a tune diagram to see how close the particle tunes are to strong coupling resonances. We track $2\times 10^{10}$ particles at initial amplitudes on a grid of $x, y$= (0 --10$\sigma_{x,y}$) for 2048 turns. Losses are defined as any particles that have amplitude growth outside the magnet half-aperture. The dynamic aperture (DA) can be estimated from the stable region in ($x,y$) amplitude space, indicated by the diffusion index $D$, which measures the tune spread between the first and last 1024 turns. A large dynamic aperture ($\approx$ the physical aperture size \cite{meot}) in a medical synchrotron is needed for high injection efficiency and long stable lifetime of therapeutic beams.
\begin{equation}{\label{eq:D}
D = \log_{10}{\sqrt{\Delta q_{x}^{2} + \Delta q_{y}^{2}}}}
\end{equation}
\vspace*{-1.0\baselineskip} 

\subsubsection*{Tolerance Studies}  
 We study the tolerance of the synchrotron with respect to tuned sextupole and octupole field gradients ($k_{2}L$, $k_{3}L$) and to potential magnet misalignment in a more practical setting, compared with an idealised simulation without placement errors. A particle survival study is performed with the same tracking parameters as for the FMA studies, this time scanning over ($k_{2}L$, $k_{3}L$) from 0 to 100$\%$ of their original values on a 12 x 12 grid. Following a method in \cite{tan}, the number of particles that survive for all initial amplitudes is plotted against each ($k_{2}L$, $k_{3}L$) on the grid to determine what combination leads to optimal particle survival ($\approx$ 100$\%$). This in turn helps to indicate how much the DA should increase. Beam dynamics simulations are then performed with the new tuned values of ($k_{2}L$, $k_{3}L$) implemented to examine the potentially improved DA. Small random magnet misalignments ($\approx$ 50-\SI{100}{\micro\meter}, equal to a 2.5$\sigma$ truncated Gaussian, comparing with studies at other synchrotron facilities \cite{Charles:2023adm}) are introduced into the lattice to gauge the effect on the DA of positioning errors that may occur in practice, compared to the influence of the higher order magnet gradients present. 
\section{Results}\label{sec:results}

\begin{figure}[!h]
\begin{center}
\begin{minipage}[!h]{0.49\linewidth}
\includegraphics[width=\linewidth]{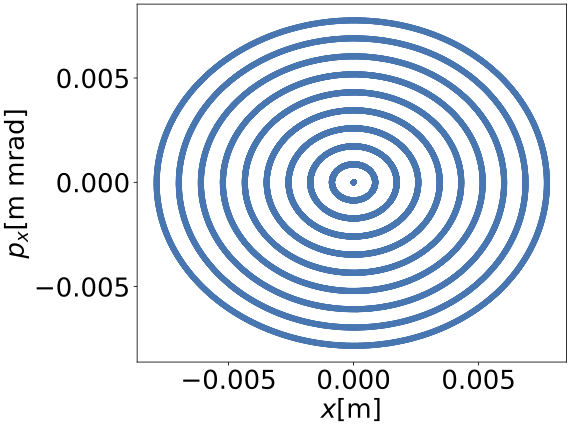}
\label{fig:phspx_og}
\end{minipage}
\begin{minipage}[!h]{0.49\linewidth}
\includegraphics[width=\linewidth]{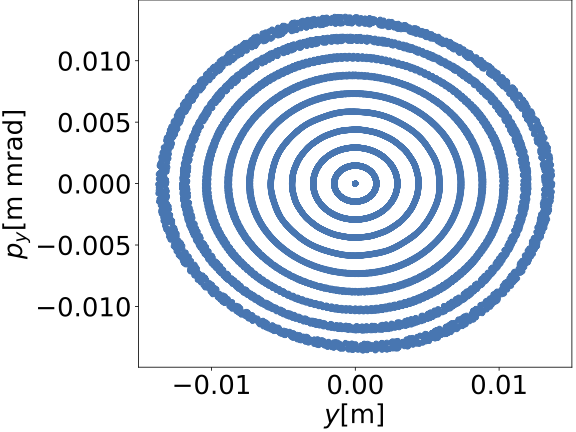}
\label{fig:phspy_og}
\end{minipage}
\begin{minipage}[!h]{0.49\linewidth}
\includegraphics[width=\linewidth]{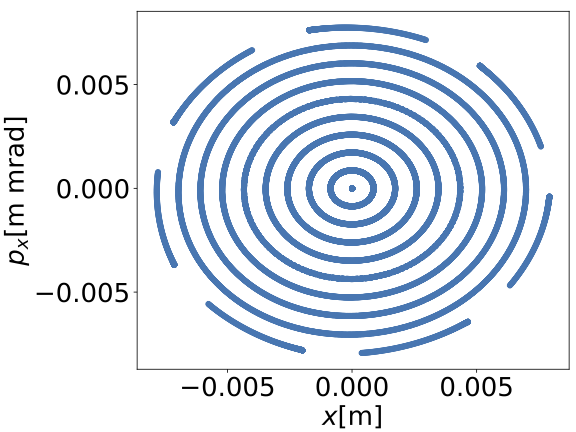}
\label{fig:phspx_multi}
\end{minipage}
\begin{minipage}[!h]{0.49\linewidth}
\includegraphics[width=\linewidth]{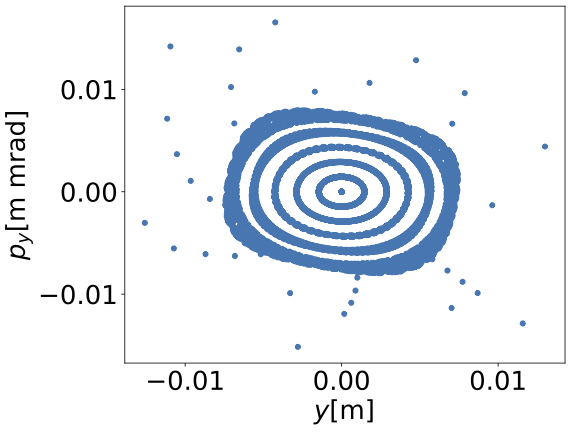}
\label{fig:phspy_multi}
\end{minipage}
\begin{minipage}[!h]{0.49\linewidth}
\includegraphics[width=\linewidth]{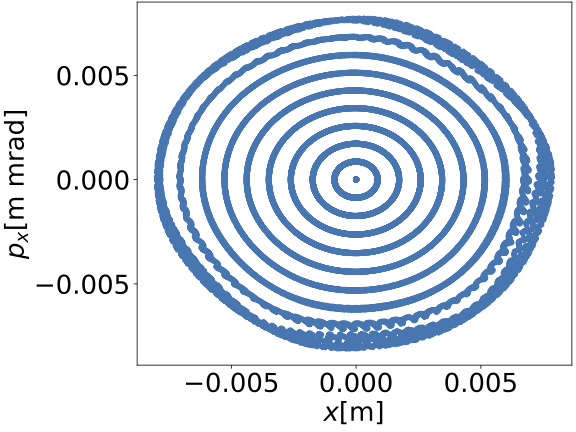}
\label{fig:phsp_x_splitmulti}
\end{minipage}
\begin{minipage}[!h]{0.49\linewidth}
\includegraphics[width=\linewidth]{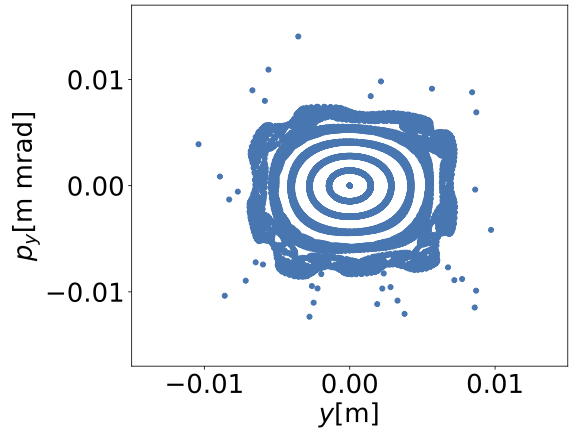}
\label{fig:phsp_y_splitmulti}
\end{minipage}
\caption{Phase space diagrams from tracking particles in $x$ (horizontal) only from 0 -- 10$\sigma_{x}$ (left) and in $y$ (vertical) only from 0 -- 10$\sigma_{y}$ (right). Top: Unmodified lattice without higher order field gradients. Middle: Lattice with integrated higher order field gradients as thin lenses between the F-D regions inside the sector bends. Bottom: Lattice with discretely sampled (non-integrated) higher order field gradients as thin lenses distributed throughout the sector bends.}
\label{fig:phspog}
\end{center}
\end{figure}
Studying the effects of incorporating the AG-CCT 3D magnetic field gradients, Fig.~\ref{fig:phspog} shows a comparison between the normalised phase space for the base lattice without the addition of any magnetic field gradients (top), the normalised phase space for the lattice with the addition of integrated field gradients (middle) and the lattice with additional thin lenses and non-integrated field gradients (bottom). For both lattices with the gradients implemented, it can be seen that their presence causes some distortion in both the horizontal and vertical phase space, expected due to the effect of higher order fields (e.g. sextupole and octupole); however no particle losses occur in the integrated gradient lattice for the starting amplitude ranges previously stated. In the non-integrated gradient lattice, it is clear that there is more distortion in both coordinates' phase spaces, with losses appearing in the vertical phase space from $\approx 5\sigma_y$.\par 

\begin{figure}[h!]
\begin{center}
\begin{minipage}[h!]{0.49\linewidth}
\includegraphics[width=\linewidth]{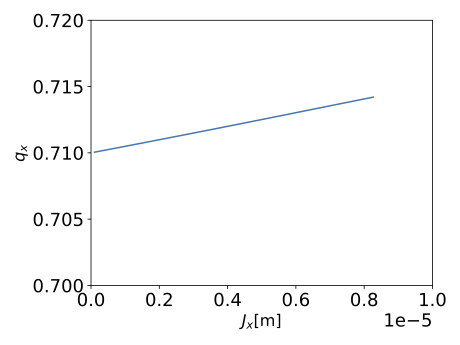}
\label{fig:xqx_multi}
\end{minipage}
\begin{minipage}[h!]{0.49\linewidth}
\includegraphics[width=\linewidth]{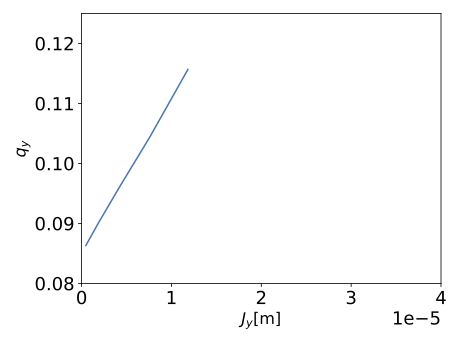}
\label{fig:yqy_multi}
\end{minipage}
\begin{minipage}[h!]{0.49\linewidth}
\includegraphics[width=\linewidth]{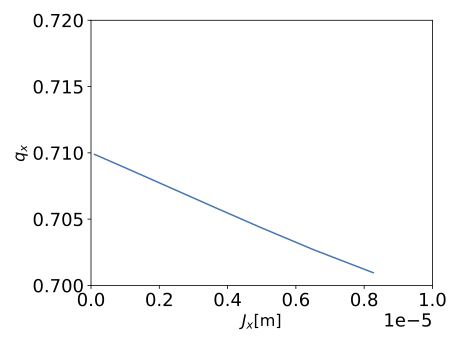}
\label{fig:xqx_split_multi}
\end{minipage}
\begin{minipage}[h!]{0.49\linewidth}
\includegraphics[width=\linewidth]{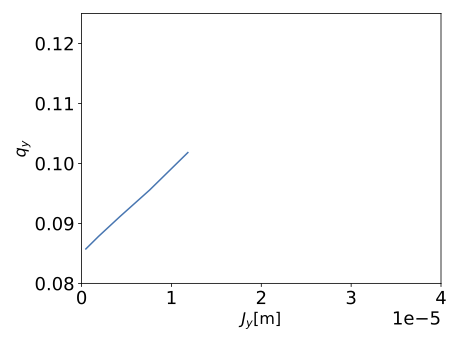}
\label{fig:yqy_split_multi}
\end{minipage}
\caption{Tune Shift with Action ($J_{x,y}$) from tracking particles in ($x,y$ = 0 -- 10$\sigma_{x,y}$). 
Top: Lattice with integrated higher order field gradients defined as thin lenses between the F-D regions inside the sector bends. Bottom: Lattice with discretely sampled (non-integrated) higher order field gradients as thin lenses distributed throughout the sector bends.}
\label{fig:tuneshift}
\end{center}
\end{figure}

The horizontal and vertical tune shift with action $J_{x,y}$ for the base lattice without any magnetic field gradients (not shown) has a constant tune with increased action, as expected. For the base lattice with integrated gradients added to the sector bends (Fig.~\ref{fig:tuneshift}, top), there are obvious shifts in tune of $\Delta q_{x}$ = 0.05 and $\Delta q_{y}$ = 0.3. Comparing to the non-integrated gradient case (Fig.~\ref{fig:tuneshift}, bottom), there is a more significant tune shift 
of $\Delta q_{x}$ = -0.1 and 
a shift of $\Delta q_{y}$ = 0.1 before losses occur for $y=5\sigma_y$. It is clearer after cross-examination of the phase space diagrams (Fig.~\ref{fig:phspog}) and tune shift with action (Fig.~\ref{fig:tuneshift}) that due to the presence of gradients, losses occur due to a large vertical tune shift: particles with a vertical tune of $q_y = 0.095$ become unstable and their vertical amplitudes grows outside the physical magnet half-aperture after excitation. 
\begin{figure}[!h]
\begin{center}
\begin{minipage}[!h]{0.9\linewidth}
\includegraphics[width=\linewidth]{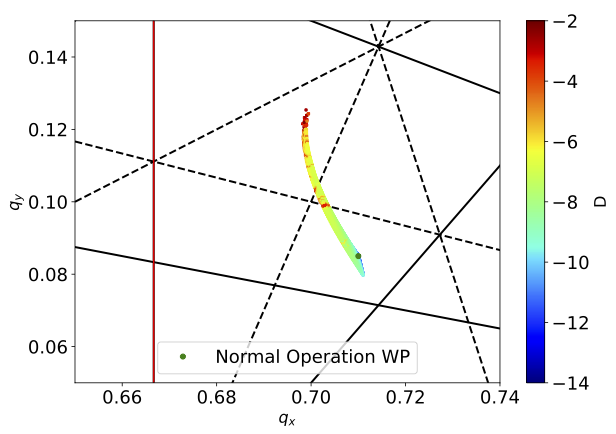}
\label{fig:fma_og_multi}
\end{minipage}
\begin{minipage}[!h]{0.9\linewidth}
\includegraphics[width=\linewidth]{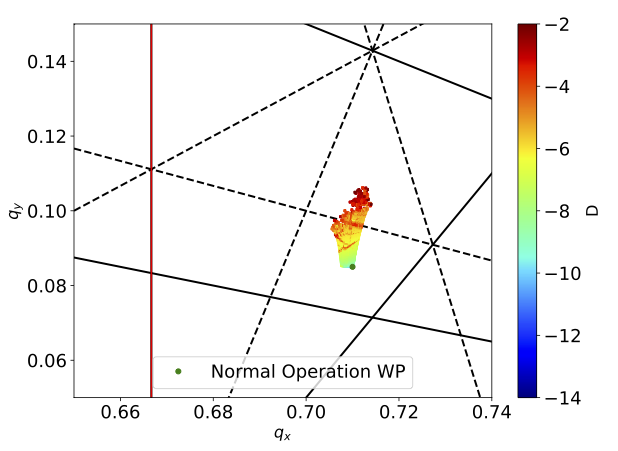}
\label{fig:fma_split_multi}
\end{minipage}
\caption{Tune space diagrams from tracking $2\times 10^{10}$ particles at initial amplitudes on a 100 x 100 grid of (-5$\sigma_{x}$ -- 5$\sigma_{x}$), ($0$ -- 10$\sigma_{y}$) for different lattices. Top: Lattice with integrated higher order field gradients as thin lenses between the F-D regions inside the sector bends. Bottom: Lattice with discretely sampled (non-integrated) higher order field gradients defined as thin lenses throughout the sector bends.}
\label{fig:fma_multi}
\end{center}
\end{figure}
\par
 Examining the calculated tunes in frequency space (Fig.~\ref{fig:fma_multi}), tune spread of the beam over time in each lattice is shown by diffusion index $D$ on the colourbar, where blue indicates more stable tunes, red equates to unstable tunes and white space indicates a particle with that particular tune has been lost during tracking. From Fig.~\ref{fig:fma_multi} (top), it can be seen that there is a large vertical tune spread over several resonances. However, there is little evidence of loss, indicating that the resonance lines crossed ($3q_{x}-q_{y}=2$, $2q_{x}+6q_{y}=2$) aren't strong enough to cause major excitation of particles at the amplitudes corresponding to those tunes. Comparing to Fig.~\ref{fig:fma_multi} (bottom), the vertical tune spread occurs near a strong 2$^\text{nd}$ order skew resonance ($3q_{x}-q_{y}=2$), which appears to be driving losses in this case. 
\begin{figure}[!h]
\begin{center}
\begin{minipage}[!h]{0.85\linewidth}
\includegraphics[width=\linewidth]{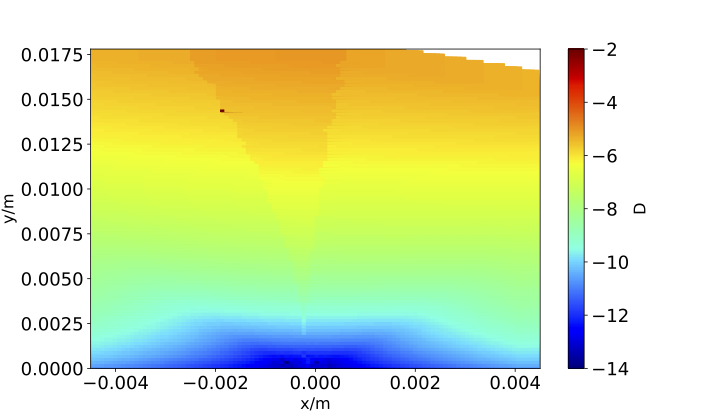}
\label{fig:da_og}
\end{minipage}
\begin{minipage}[!h]{0.85\linewidth}
\includegraphics[width=\linewidth]{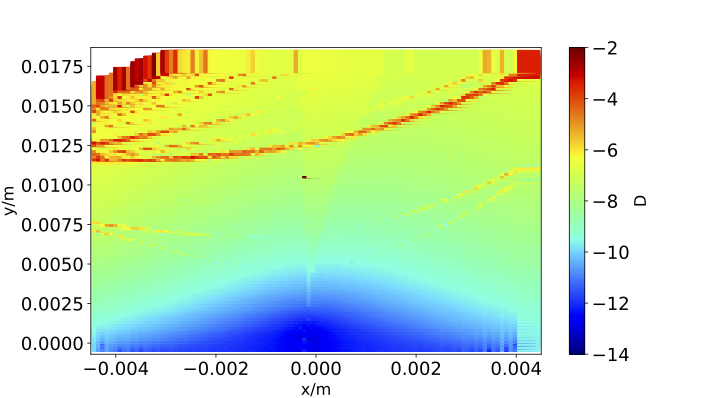}
\label{fig:da_og_multi}
\end{minipage}
\begin{minipage}[!h]{0.85\linewidth}
\includegraphics[width=\linewidth]{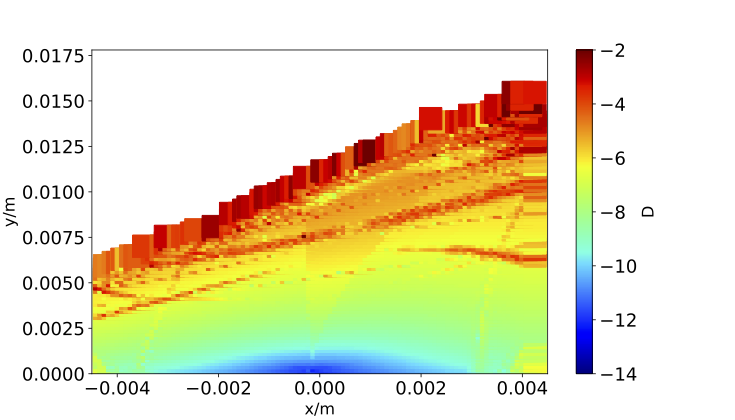}
\label{fig:da_split_multi}
\end{minipage}
\caption{Initial amplitude space diagrams constructed from tracking $2\times 10^{10}$ particles at initial amplitudes on a 100 x 100 grid of (-5$\sigma_{x}$ -- 5$\sigma_{x}$), (0 -- 10$\sigma_{y}$). Top: Base lattice without the presence of any magnetic field gradients. Middle: Lattice with integrated higher order field gradients as thin lenses between the F-D regions inside the sector bends. Bottom: Lattice with discretely sampled (non-integrated) higher order field gradients as thin lenses distributed throughout the sector bends.}
\label{fig:da}
\end{center}
\end{figure}
\par
Viewing the same tracking data in ($x,y$) amplitude space (Fig.~\ref{fig:da}, middle and bottom), we can estimate the DA from the stable regions in the diagrams and evaluate any shrinkage due to the gradients by comparing with the base lattice without gradients (Fig.~\ref{fig:da}, top). Similar to Fig.~\ref{fig:fma_multi}, the initial amplitudes of the particles are compared to their tune spread after tracking. Using the colourbar, the area of the stable region (indicated from blue to yellow) allows estimation of the dynamic aperture. Comparing the top and middle graphs in Fig.~\ref{fig:da}, it can be seen that the presence of integrated gradients is well-tolerated by the synchrotron. The DA is reduced only slightly in the vertical plane from $|y|$=0.012 to 0.010 compared to that of the base lattice. However, comparing to the non-integrated gradient representation results (Fig.~\ref{fig:da}) (bottom)), there is a dramatic decrease of the vertical DA by 75$\%$ ($|y|$=0.003, or $\approx 2\sigma_y$). This result justifies our representation of magnetic fields for beam dynamics with discretely sampled field gradients along the entire CCT magnet curvature. Otherwise, higher order field detail due to curvature may be missed using an integrated gradient approach, and a much larger DA might be falsely achieved. Though the presence of the curved field gradients reduces the DA by 75$\%$, the simulated synchrotron is able to store particles with up to 10$\sigma_{x,y}$  initial amplitude offset: resonances do not appear to cause significant beam loss in this amplitude range. There is no set design limit on the DA, however we wish to maximise it as much as possible for optimal performance, aiming towards $10 \sigma_{x,y}$ (the magnet half-aperture).  \par
\begin{figure}[h!] 
\begin{center}
\includegraphics[width=.9\linewidth]{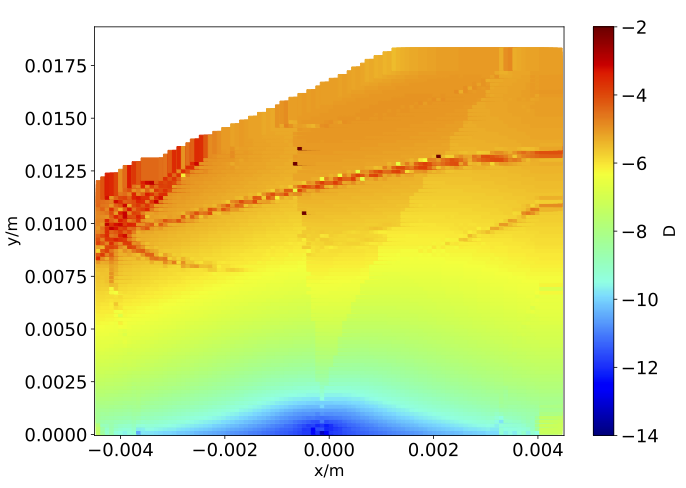} 
\caption{Initial amplitude space in the non-integrated gradient lattice after tuning higher order gradients $k_{2}L$, $k_{3}L$.} 
\label{fig:tolstudyda}
\end{center}
\end{figure}
\par
We have demonstrated that the model carbon therapy synchrotron lattice appears able to store $\epsilon_{xN,yN}$ = \SI{0.7}{\text{$\mu$}\meter},\SI{1.0}{\text{$\mu$}\meter} carbon ions for 2048 turns. However, for ultimate performance and beam stability, the DA may need further improvement depending upon the operational needs of a treatment facility. We have therefore further investigated the intentional tuning of higher order field gradients in order to improve the DA, which may be required for the next iteration of design of the AG-CCT magnet. Fig.~\ref{fig:tolstudy} shows that ($k_{2}L$, $k_{3}L$) must be tuned to 0.56 and 0.25 times their respective values to yield the highest particle survival and potentially increase the DA. Using these tuned 
gradients, particles with the same tracking parameters as in Section~\ref{sec:beamdynamics} are tracked to determine the new DA, as shown in Fig.~\ref{fig:tolstudyda}. The vertical DA improves from $|y|$=\SI{0.3}{\meter} to $|y|$=\SI{0.6}{\meter}, or $\approx$ 50$\%$ after tuning. This means that in the presence of unwanted higher order gradients arising from the CCT curvature, the DA of the NIMMS synchrotron can be mostly recovered with adjustment of the higher order field gradients.  

\begin{figure}[h!]
\begin{center} 
\includegraphics[width=\linewidth]{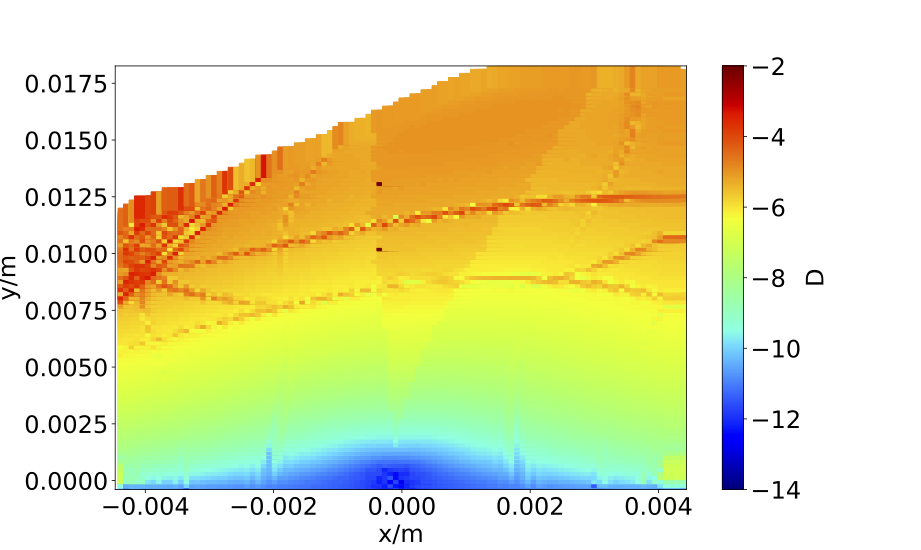}
\caption{Intial amplitude space in the lattice with discretely sampled, tuned gradients after misalignments of \SI{100}{\micro\meter} in the sector bends, \SI{50}{\micro\meter} in the quadrupoles and sextupoles.}
\label{fig:misalign}
\end{center}
\end{figure}
\par
Further adding to the realism of the simulation, we also investigate the DA when small misalignments (as earlier described in Section~\ref{sec:beamdynamics}) are introduced to both the base lattice and the lattice with discretely sampled, tuned gradients. The effect on the DA of the misalignments is practically negligible in both the base lattice and in the lattice with tuned gradients implemented. The DA is much more strongly influenced by the presence of magnet field gradients than the misalignments, as can be seen by comparing Figs.~\ref{fig:tolstudyda}, {\ref{fig:misalign}}.\par 
The key results from this study show distortion arising in the phase space due to curved magnetic field gradients, leading to a reduction in the vertical DA by $75\%$ using discretely sampled gradients in a thin lens representation, compared to a $17\%$ reduction using an integrated gradient representation. By using integrated gradients, one may be led incorrectly to assume the region of stability in the lattice to be much larger than reality: while this may be a sufficient representation for magnets within larger accelerators \cite{esrf,fcc,storagering}, we have demonstrated this is not the case here. The DA of the synchrotron can be mostly recovered by tuning sextupole and octupole gradients to (0.56, 0.25) times their initial values, which would require additional magnet layers in practice \cite{holger}. 
We found that there was no significant impact on the DA from small sector bend misalignments that may occur in practice.
\section{Conclusion}

The goals of this paper were to analyse a strongly curved, superconducting, combined-function CCT magnet in context of a compact synchrotron for heavy ion therapy, and to assess the impact of the curvature on the field content and the beam dynamics. This required the accurate analysis of non-linear fields due to curvature and a simulation pipeline sufficient to model the compact accelerator, which is necessary to determine the suitability of the magnet in this application, but has not been previously performed. 

To address these goals, we devised a tool to sample fields from simulated magnetic fieldmaps in OPERA 3D and calculate magnetic field gradients, as multipole expansion is no longer mathematically valid for field analysis. 
Our choice of the field gradient representation in the curved coordinate system was found to be justified, overcoming the mismatch between curved gradients and multipoles. An extension of the curved field gradient calculation to 3D was performed for fully representative field modelling in the curved coordinate system of the AG-CCT. 
\par The 3D normal and skew gradients were implemented in the synchrotron lattice in MAD-X for beam dynamics simulations using the in-built PTC module to investigate the gradients' effects on the compact synchrotron operation. It was shown that the commonly used integrated gradient representation did not reflect accurately the phase space distortion and particle losses on resonances due to higher order fields. After improvement of the representation using discrete field sampling along the magnet length, we discovered significantly larger tune shifts due to additional finite field detail, leading to a reduction in the vertical dynamic aperture by $75\%$ After tuning sextupole and octupole gradients by means of a particle survival study, the vertical dynamic aperture improved to $50\%$. 
\par Our results show a considerable reduction in the DA compared to the initial particle amplitudes
tracked within the physical half-aperture, however we have assessed in this study that this is tolerable for stable operation of the synchrotron. We further checked the impact of realistic magnet misalignments on beam stability, finding little further reduction in DA compared to previous results, concluding that for the range of misalignments introduced, the nonlinear fields arising from magnet curvature have the dominant impact on DA. \par 
In conclusion, our studies assess that the presence of curved magnetic field gradients do not severely restrict the operation of the medical synchrotron studied. However, the gradients are important to assess for other compact accelerator scenarios or in the case the achieved DA is insufficient. In this case, the CCT magnet design must be adjusted accordingly by e.g. adding additional coil layers for tuning. In its current iteration, we assess that the synchrotron based on curved AG-CCT magnets is capable of achieving the design goals laid out in the introduction of this paper, with the potential to enable future heavy ion therapy facilities for improved patient access and clinical outcomes.
\section*{Acknowledgements}
This work has received funding from the Engineering and Physical Sciences Research Council, UK. The authors would like to thank our collaborators within the NIMMS and HITRI+ groups for providing valuable feedback and guidance during this project, and to Mikko Karpinnen for provision of the CF-CT OPERA models. Additional thanks goes to our collaborators at Lawrence Berkeley National Laboratory for provision of the AG-CCT OPERA model. A final note of thanks to Ben Pine at Dassault Syst\`{e}mes for valuable support with OPERA.
\nocite{*}

\bibliography{Effects_of_curved_magnetic_fields_on_beam_stability}
\appendix
\section{\label{app:boundary}Boundary Conditions}

\begin{equation}\label{eq:boundary}
{V(r,\theta) = \begin{cases}
  \phi(\theta + L)  & \text{for }r > 0\\    
  < \infty  & \text{for } r = 0 \\
  V(\theta)& \text{for }r = r_{0}\\ 
\end{cases}}
\end{equation}
$L$ = an integer multiple of $2\pi$, $r$ is the radial distance measured from the nominal beam trajectory, $\theta$ is the angle subtended between the magnet and beam $r_{0}$ is the reference radius.

\section{\label{app:altsoln}Alternative Solution: Field Derivatives in 2D Straight and Curved Systems}
In Cartesian coordinates, this can be chosen along the vertical component $y = 0$ and $\textbf{B}$$(x,y) = B_y(x,y)$, where transverse components ($x,y$) are centered on the nominal trajectory and $B_x = -B_x = B_z=-B_z = 0$. This has the form 
\begin{equation}\label{eq:taylor_grad}
   B_{y}(x\mid_{y=0}) = B_{y} + \frac{\text{d}B_{y}}{\text{d}x}x +\frac{1}{2!} \frac{\text{d}^2{B_y}}{\text{d}x^2}x^2+\frac{1}{3!} \frac{\text{d}^3{B_y}}{\text{d}x^3}x^3+...
\end{equation}
Where all $B_{y}$ are evaluated at ($x,y)=0$ and the first term represents the dipole field, the second term represents the quadrupole field, and so on. By comparison, the multipole expansion of the field in straight Cartesian coordinates is
\begin{equation}\label{eq:taylor_multi}
    B_y(x)\mid_{y=0} = B_0 + \frac{\text{1}}{{r}_{0}}B_{1}x +\frac{1}{2!} \frac{\text{1}}{{r}_{0}^2}B_{2}x^{2}+\frac{1}{3!} \frac{{1}}{{r}_{0}^3}B_{3}x^{3}+...
\end{equation}
The multipole coefficients are evaluated at $r_{0}$ as dictated by \ref{eq:boundary}; the normalisation factor requires a reference radius to be defined, due to the increase of the field strength as a polynomial function of the radius, whose order is $(n-1)$. It can be seen that the two representations are equivalent, aside from a normalisation factor ${r_0^{-n}}$ in \ref{eq:taylor_multi}.\par
On the magnet midplane, the components of \textbf{B} follow from Ref.~\cite{slac75} as
\begin{equation}\label{eq:curvedBx}
B_{x}(x,y,s)=\frac{\partial \phi(x,y,s)}{\partial x}=c_{11} y + c_{12} xy + ...
\end{equation}
\begin{equation}\label{eq:curvedBy}
\begin{split}
    B_{y}(x,y,s) & = \frac{\partial \phi(x,y,s)}{\partial y}=\\ & c_{10}  + c_{11} x + \frac{1}{2!}c_{12}x^{2} + \frac{1}{2!}c_{30}y^{2} + ...\\ & c_{30}  = [h_{s}c_{11}+c_{12}]
\end{split}
\end{equation}
\begin{equation}\label{eq:curvedBs}
B_{s}(x,y,s)=\frac{1}{h_s}\frac{\partial \phi(x,y,s)}{\partial s}=\frac{1}{h_s} \left[c^\prime_{10} y + c^\prime_{11} xy + ... \right]
\end{equation}
where scale factor $h_s$ = (1 + $\frac{x}{\rho}$) \cite{Lee}.
Here the prime denotes the total derivative with respect to $s$. 
In a rectangular coordinate system, $\rho \rightarrow \infty$ and $h_s$ = 1, which makes $y$=0 and reduces \ref{eq:curvedBy} back to \ref{eq:taylor_grad} as before. 
\section{\label{app:tabs}Calculating Gradients and Multipoles for Different Coordinate Systems}
\begin{figure}[!h]
\begin{center}
\begin{minipage}[!h]{0.6\linewidth}
\includegraphics[width=\linewidth]{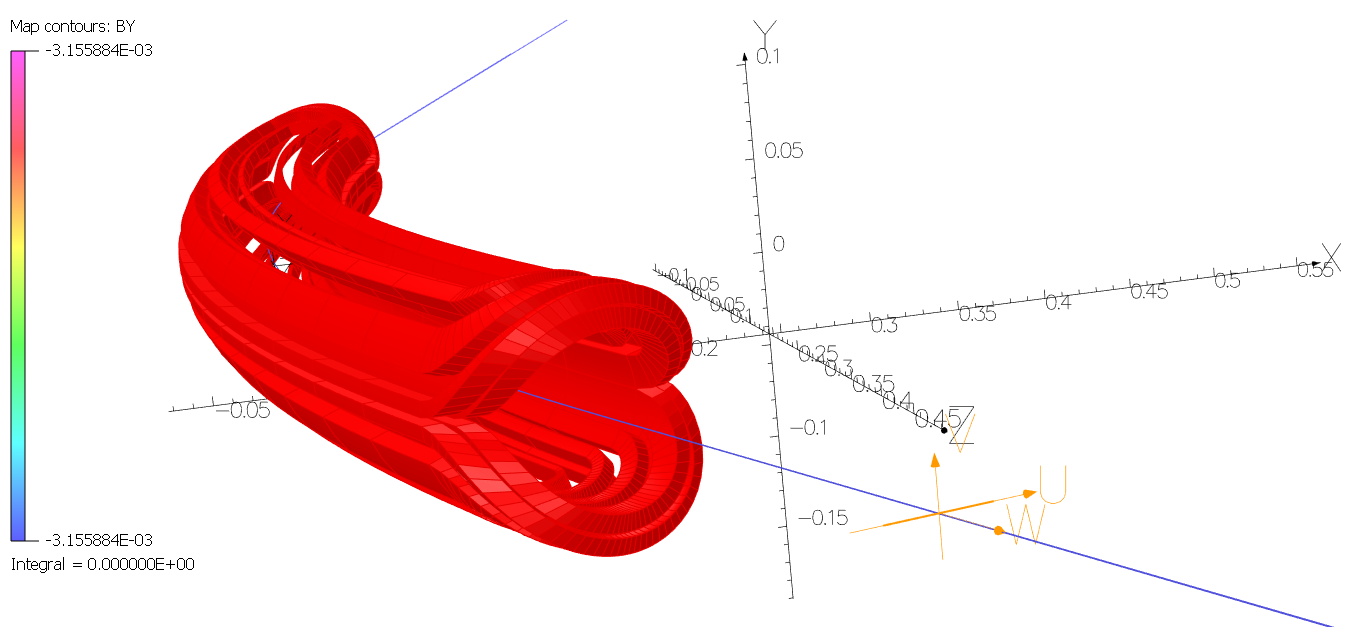}
\label{fig:curvedsampleline}
\end{minipage}
\begin{minipage}[!h]{0.6\linewidth}
\includegraphics[width=\linewidth]{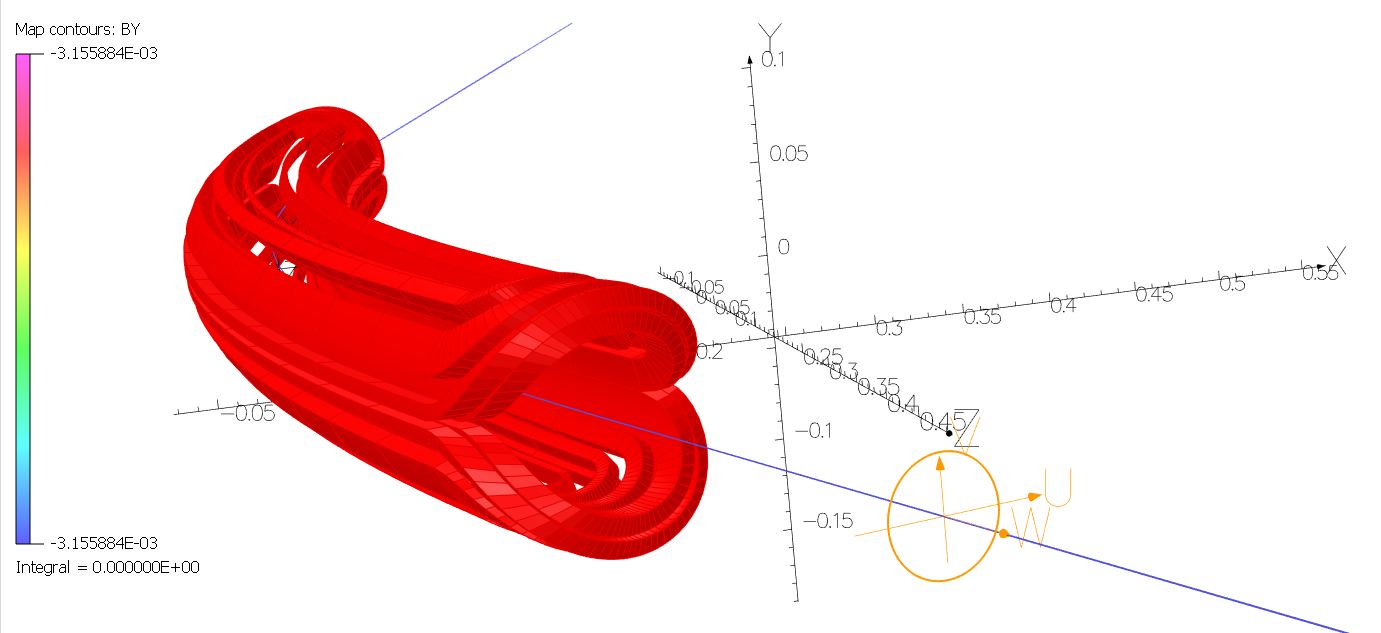} 
\label{fig:curvedsamplecircle}
\end{minipage}
\caption{An ideal trajectory through a curved CT magnetic field, with Top: a line sampled along $y$=0 perpendicular to the path and Bottom: a circle defined at a radius = \SI{0.02}{\meter} (2/3 the aperture radius) sampled perpendicular to the path. The coil and path are shown with the local coordinate systems defined for both depictions of sampling.}
\label{fig:sampling}
\end{center}
\end{figure}
\begin{table}[!hbt]
\centering
\caption{2D straight system (ideal magnet): A comparison between arbitrary initial input gradients, the gradients converted from the coefficient terms of a polynomial fitted to main field $B_{y}(y=0)$, and the gradients converted from the multipole coefficients calculated from a DFT on the radial component of the main field $B_{r}(r,\theta)$ for a few orders of $n =$(0,...,3).} 
\begin{ruledtabular}
\begin{tabular}{llll}
\centering
Order(n) & Input $g_{n}$ & Polynomial Fit $g_{n}$                 & Fourier Analysis $g_{n}$          \\ 
0        & $\SI{3.0}{\tesla}$            & $\SI{3.0} {\tesla}$                         & $\SI{3.0} {\tesla}$              \\ 
1        & $\SI{1.4}{\tesla/\meter}$          & $\SI{1.4} {\tesla/\meter}$                 & $\SI{1.4} {\tesla/\meter}$               \\ 
2        & $\SI{16}{\tesla/\meter^{2}}$               & $\SI{16}{\tesla/\meter^{2}}$             & $\SI{16}{\tesla/\meter^{2}}$    \\ 
3        & $\SI{33}{\tesla/\meter^{3}}$              & $\SI{33}{\tesla/\meter^{3}}$ & $\SI{33}{\tesla/\meter^{3}}$   
\label{tab:ideal}
\end{tabular}
\end{ruledtabular}
\end{table}

\begin{table}[!htb]
\centering
\caption{2D straight system (realistic magnet): A comparison of integrated gradient harmonics converted from the coefficients of a polynomial fitted to main field $B_{y}$ and integrated multipole harmonics calculated from sampling the radial field $B_r$ around a circle of radius ($r_0 = \SI{0.02}{\meter}$) for a few orders of $n =$(0,...,3), between $z$ = -0.4 and \SI{0.4}{\meter} (the magnet body). Each harmonic $b_n$ is given in terms of the main field $B_\text{main}$: $b_n$=$B_{n}$/$B_\text{main}*10^{-4}$}
\begin{ruledtabular}
\begin{tabular}{llll}
Order(n) & Polynomial Fit $b_{n}(x)$ & Fourier Analysis $b_{n}(x)$                  & Unit         \\ \hline
0        & ${2.77}$            & ${2.77} $                         & Tm               \\ 
1        & ${-331.91}$          & ${-331.88}$                 & $10^{-4}$                \\ 
2        & ${15.68}$               & ${15.56}$             & $10^{-4}$    \\ 
3        & ${0.585}$              & ${0.109}$ & $10^{-4}$   \\ 
\end{tabular}
\end{ruledtabular}
\label{tab:integratedbody}
\end{table}

\begin{table}[!htb]
\centering
\caption{2D curved system: A comparison of integrated gradient harmonics converted from the coefficients of a polynomial fitted to main field $B_{y}$ and integrated multipole harmonics calculated from sampling the radial field $B_r$ around a circle of radius ($r_0 = \SI{0.02}{\meter}$) for a few orders of $n =$(0,...,3), between $z$ = -0.8 and \SI{0.8}{\meter} (the entire magnet length). As before, each $b_n$=$B_{n}$/$B_\text{main}*10^{-4}$.} 
\begin{ruledtabular}
\begin{tabular}{llll}
Order(n) & Polynomial Fit $b_{n}(x)$ & Fourier Analysis $b_{n}(x)$                  & Unit        \\ \hline
0        & ${3.39}$            & ${3.39} $                         & Tm               \\ 
1        & ${310.0}$          & ${264.18}$                 & $10^{-4}$                \\ 
2        & ${5.16}$               & ${1.50}$             & $10^{-4}$    \\ 
3        & ${16.04}$              & ${1.26}$ & $10^{-4}$   \\ 
\end{tabular}
\end{ruledtabular}
\label{tab:curvedcomp}
\end{table}

\newpage
\section{\label{app:survival} Particle Survival Study}

\begin{figure}[h!] 
\begin{center}
\includegraphics[width=.7\linewidth]{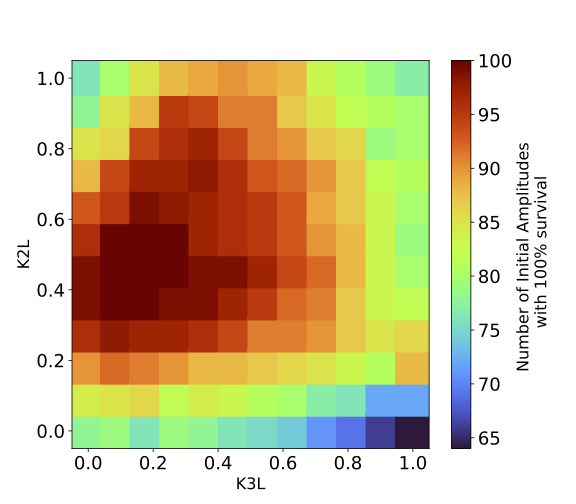} 
\caption{Particle survival rate of 100$\%$ corresponding to number of initial amplitudes, tracked from (5$\sigma_{x}$ -- 5$\sigma_{x}$), (0 --10$\sigma_{y}$) in a 10 x 10 grid in the non-integrated gradient lattice. Varying ($k_{2}L$ = sextupole, $k_{3}L$ = octupole) between 0 and 100$\%$ of their previous values. 
}
\label{fig:tolstudy}
\end{center}
\end{figure}
\newpage
\end{document}